\documentclass[fleqn,usenatbib]{mnras}

\usepackage[T1]{fontenc}

\DeclareRobustCommand{\VAN}[3]{#2}
\let\VANthebibliography\thebibliography
\def\thebibliography{\DeclareRobustCommand{\VAN}[3]{##3}\VANthebibliography}

\usepackage{graphicx}	
\usepackage{amsmath}	

\renewcommand{\vec}[1]{ {\mathbf #1} }

\newcommand{\Eq}{{Equation}}

\newcommand{\Fig}{{Figure}}
\newcommand{\Figs}{{Figures}}

\title[CME modeling]{MHD Modeling of the Near-Sun Evolution of Coronal Mass Ejection Initiated from a Sheared Arcade}

\author[Cai et al.]{Jinnan Cai, Ling Zhang,
  Chaowei~Jiang\thanks{E-mail: chaowei@hit.edu.cn (CWJ)}, Kuo Yan,
  Xueshang Feng, Pingbing Zuo, Yi Wang
  \\
  Shenzhen Key Laboratory of Numerical Prediction for Space Storm,
  School of Aerospace Science,\\ Harbin Institute of Technology,
  Shenzhen 518055, China}
\date{Accepted XXX. Received YYY; in original form ZZZ}

\pubyear{2025}

\begin{document}
\label{firstpage}
\pagerange{\pageref{firstpage}--\pageref{lastpage}}
\maketitle

\begin{abstract}
  Coronal mass ejections (CMEs) are phenomena in which the Sun
  suddenly releases a mass of energy and magnetized plasma,
  potentially leading to adverse space weather. Numerical simulation
  provides an important avenue for comprehensively understanding the
  structure and mechanism of CMEs. Here we present a global-corona MHD
  simulation of a CME originating from sheared magnetic arcade and its
  interaction with the near-Sun solar wind. Our simulation
  encompasses the pre-CME phase with gradual
  accumulation of free magnetic energy (and building up of a current
  sheet within the sheared arcade) as driven by the photospheric
  shearing motion, the initiation of CME as magnetic reconnection
  commences at the current sheet, and its subsequent evolution and
  propagation to around $0.1$~AU. A twisted magnetic flux rope (MFR),
  as the main body of the CME, is created by the continuous
  reconnection during the eruption. By interacting with the ambient
  field, the MFR experiences both rotation and deflection during the
  evolution. The CME exhibits a typical three-part structure, namely a
  bright core, a dark cavity and a bright front. The bright core is
  mainly located at the lower part of the MFR, where plasma is rapidly
  pumped in by the high-speed reconnection outflow. The dark cavity
  contains both outer layer of the MFR and its overlying field that
  expands rapidly as the whole magnetic structure moves out. The
  bright front is formed due to compression of plasma ahead of the
  fast-moving magnetic structure. Future data-driven modeling of CME will
  be built upon this simulation with real observations used for the bottom boundary conditions.
\end{abstract}


\section{Introduction}
\label{sect:intro}

Coronal mass ejections (CMEs) are the largest scale solar activities,
characterized by a mass of magnetized plasma being ejected from the
solar atmosphere into interplanetary space, potentially impacting
Earth and causing harmful space weather effects. Since they were first
observed from space with the coronagraph onboard NASA's Seventh
Orbiting Solar Observatory (OSO-7) on 14 December 1971, CMEs have
garnered widespread attention, leading to extensive research across
observations, theoretical analysis, and numerical simulations. The
recent decades of research have provided valuable models and
explanations regarding the precursor structures, triggering
mechanisms, and propagation evolution of
CMEs~\citep{
  gopalswamyGlobalPictureCMEs2004,
  forbesCMETheoryModels2006,
  chenCoronalMassEjections2011,
  webbCoronalMassEjections2012,
  kleimann4pModelsCMEs2012,
  manchester_PhysicalProcessesCME_2017,
  chenPhysicsEruptingSolar2017,
  luhmannICMEEvolutionInner2020,
  jiang_FundamentalMechanismSolar_2021,
  zhangEarthaffectingSolarTransients2021, jiangFundamentalTheoryRealistic2024}.

Nevertheless, due to the limitations of current observations, we are
still far away from a comprehensive understanding of the 3D structure
and evolution of CMEs and the underlying mechanisms~\citep{lugazImportanceFundamentalResearch2023, torokLearnWalkYou2023, temmerCMEPropagationHeliosphere2023}. For example,
primarily due to the difficulty in obtaining the 3D magnetic field
structure in the low corona, the initiation mechanism of CMEs is a
topic of controversy for many years~\citep{forbesCMETheoryModels2006,
  chenCoronalMassEjections2011, aulanierPhysicalMechanismsThat2013,
  schmiederFlareCMEModelsObservational2015}. Some argued that CMEs are
caused by the loss of equilibrium of pre-existing twisted magnetic
flux ropes (MFRs) due to some kind of ideal MHD
instability~\citep[e.g.,][]{amari_CoronalMassEjection_2003,
  torokConfinedEjectiveEruptions2005,
  aulanier_FormationTorusunstableFlux_2010}, while others emphasized
the key role of magnetic reconnection due to complex magnetic topology
in initiating the
eruption~\citep[e.g.,][]{antiochos_ModelSolarCoronal_1999,
  chenEmergingFluxTrigger2000, kusanoMagneticFieldStructures2012}. In
observations, it is difficult to distinguish the specific mechanisms
\citep{howard_InnerHeliosphericFlux_2012a,
  kumar_MultiwavelengthObservationsEruptive_2013,
  cheng_TrackingEvolutionCoherent_2014}. Another unresolved issue is
the nature of the classic three-part structure of many CMEs as seen in
coronagraph: a bright core, a dark cavity and a bright leading edge
\citep{illing_ObservationCoronalTransient_1985a}. A conventional view
is that the bright front is formed by plasma pileup along the outer
edge of the MFR, the cavity corresponds to the main body of the MFR
and the bright core is the erupted prominence (or filament) at the
dipped portion of the MFR
\citep[e.g.][]{bothmer_StructureOriginMagnetic_1998,
  forbes_ReviewGenesisCoronal_2000},
as early studies suggested that CMEs were more closely related to
filaments than flares~\citep{gosling_SpeedsCoronalMass_1976,
  joselyn_DisappearingSolarFilaments_1981}.  However, subsequent
statistical researches found that only a very small portion of CMEs
are associated with filament eruptions
\citep{lepri_DirectObservationalEvidence_2010,
  wood_ImagingProminenceEruptions_2016}, prompting
\citet{howard_ChallengingContemporaryViews_2016} to question whether
the bright cores of CMEs are not filaments but rather the natural
result of MFR propagation or the visual effect presented by 3D
extended MFRs. Later, it is demonstrated that
many CMEs unrelated to filaments also exhibit a three-part structure~\citep{songThreepartStructureFilamentunrelated2017},
and a different view is proposed that the core and front correspond to
the MFR plasma and plasma pileup along the coronal loops,
respectively, while the cavity is either a part of the MFR, or a
low-density zone between the front and the
MFR~\citep{songStructureCoronalMass2023, songNatureDarkCavity2025}.


Numerical simulation has long been an important way of investigating
the initiation mechanism, structure and evolution of
CMEs~\citep[e.g.,][]{mikicDisruptionCoronalMagnetic1994,
  grothGlobalThreedimensionalMHD2000,
  manchesterThreedimensionalMHDSimulation2004,
  vanderholstModellingSolarWind2005,
  rileyModelingInterplanetaryCoronal2006,
  kataokaThreedimensionalMHDModeling2009,
  lugazNUMERICALINVESTIGATIONCORONAL2011,
  lionelloMagnetohydrodynamicSimulationsInterplanetary2013,
  zhouUsingMHDSimulation2014, shenTurnSuperelasticCollision2016,
  jinDATACONSTRAINEDCORONALMASS2017a,
  torokSuntoEarthMHDSimulation2018, yangNumericalMHDSimulations2021,
  koehnSuccessiveInteractingCoronal2022,
  meiNumericalSimulationLeading2023, guoDependenceCoronalMass2024,
  linanCoronalMassEjection2024}. Recently, with a high-resolution MHD
simulation, \citet{jiang_FundamentalMechanismSolar_2021} established a
fundamental mechanism for CME initiations, in which an internal
current sheet forms gradually within a continuously sheared magnetic
arcade as driven by photospheric motions and fast reconnection at this
current sheet initiating the eruption. However, their simulation
region is limited to a local Cartesian box, which only approximates
the corona of active region size, while the evolution of a CME is
often a global behavior due to its fast expansion. Furthermore, the
interaction of CME with the background solar wind is also an important
factor in shaping the CME structure. In this paper, we extend
\citet{jiang_FundamentalMechanismSolar_2021}'s simulation to the
global corona with a polytropic solar wind background from the solar
surface to around $0.1$~AU. This advanced simulation allows us to the
study the initiation and near-Sun evolution of CMEs.



\begin{figure*}
    \centering
    \includegraphics[width=0.8\textwidth]{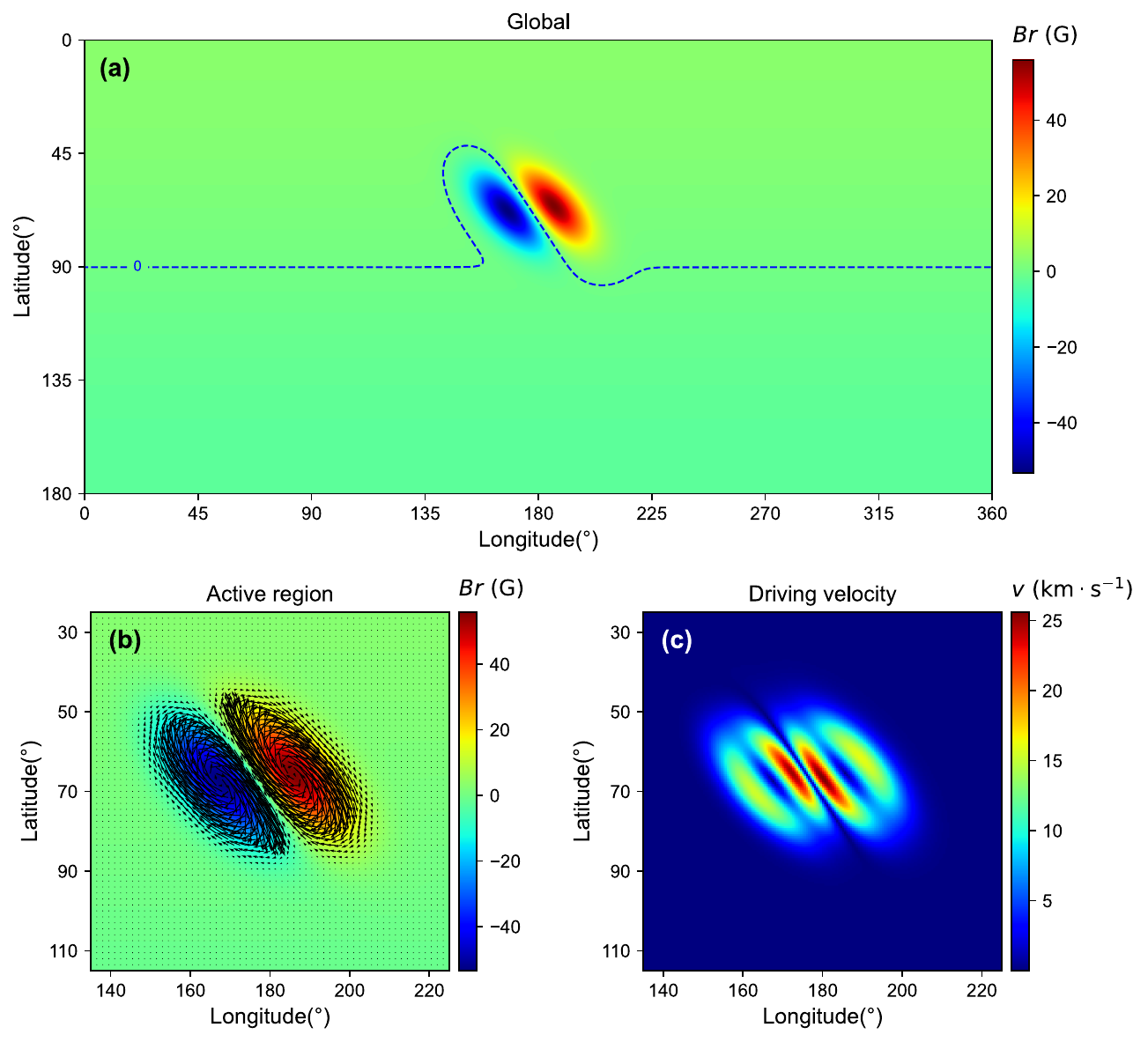}
    \caption{Map of magnetic flux density on the solar surface. (a)
      The global map. The PIL is shown by the dashed curve. (b) The
      active region; the arrows show the surface driving flow. (c) The
      driving velocity.}
    \label{fig:initialCondition}
\end{figure*}

\begin{figure*}
    \centering
    \includegraphics[width=\textwidth]{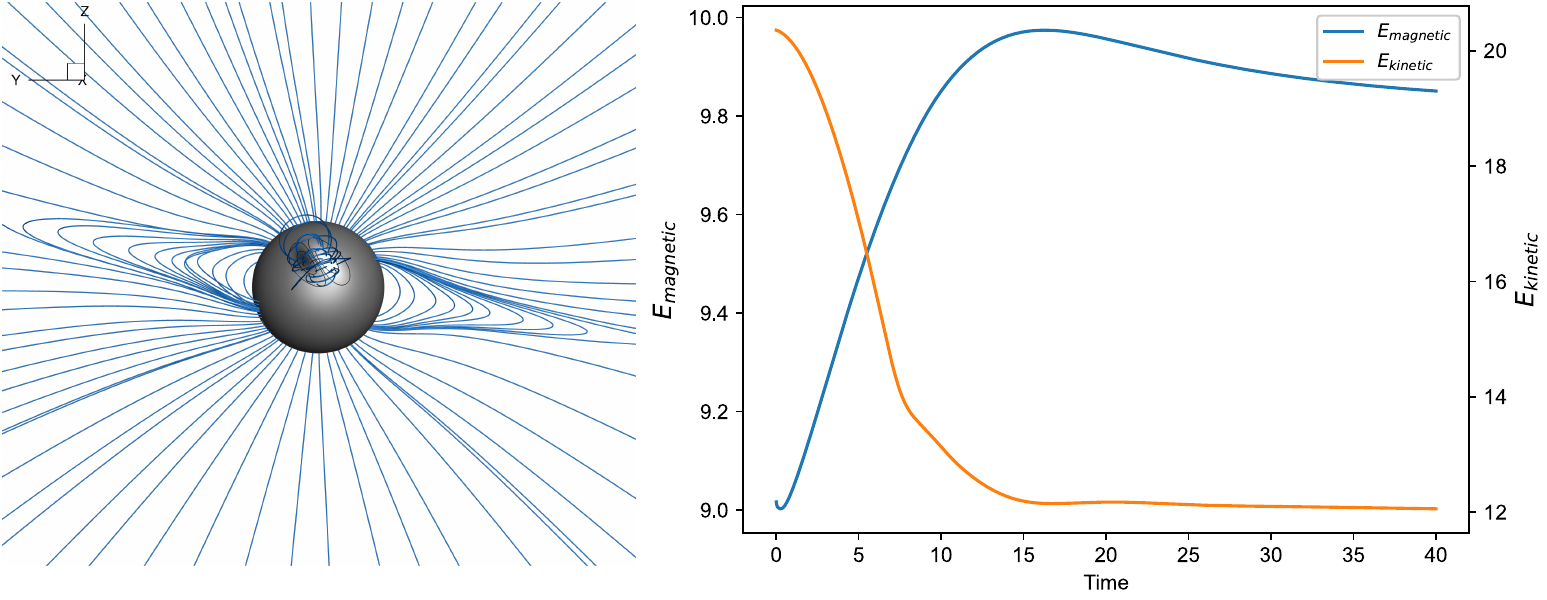}
    \caption{Simulation of the steady-state solar wind. In the left panel, the sphere
      shows the solar surface with contour of magnetic flux density
      $B_r$. The lines are magnetic field lines. The right panel shows temporal
      evolution of the magnetic and kinetic energies in the relaxation
      process.}
    \label{fig:backgroundSolarWind}
\end{figure*}

\section{Numerical Model}

\subsection{The control equations}
We numerically solve the 3D MHD equations in 3D using the
AMR--CESE--MHD
code~\citep{jiang_AMRSimulationsMagnetohydrodynamic_2010}. The MHD
equations are given as
\begin{eqnarray}
  \frac{\partial \rho}{\partial t} + \nabla \cdot (\rho \vec {v}) = 0, \nonumber
  \\
  \rho \frac{D \vec {v}}{Dt} = -\vec {\nabla p + \vec J \times \vec B }+ \rho \vec {g} + \nabla
  \cdot (\nu \rho \nabla \vec {v}), \nonumber
  \\
  \frac{\partial \vec {B}}{\partial t} =
  \nabla \times ( \vec {v \times \vec B} ), \nonumber
  \\
  \frac{\partial T}{\partial t} + \nabla \cdot (T \vec {v})=
  (2 - \gamma)T \nabla \cdot \vec {v}.
\end{eqnarray}
Here $\vec {v}$ represents the velocity,
$\vec {J=\nabla \times \vec B}/\mu_0$ (with $\mu_0$ denotes the
magnetic permeability in a vacuum) is the current density, $\vec {g}$
refers to the gravitational acceleration exerted by the Sun, $T$ is
the temperature, and $\gamma$ is the adiabatic index, which is given
as $\gamma = 1.05$ to approximate a near-isothermal process.

In the code, all the variables are normalized by typical values in the
corona. The values are, respectively, $L_{s} = 6.96\times 10^{2}$~Mm
(i.e., solar radius) for length,
$\rho_{s} = 1.67\times 10^{-15}$~g~cm$^{-3}$ for density,
$T_{s} = 10^{6}$~K for temperature, $p_{s} = \rho_{s}RT_{s}$ (where
$R$ is the gas constant) for pressure,
$v_{s} = \sqrt{p_{s}/\rho_{s}} = 1.28\times 10^{2}$~km~s$^{-1}$ for
velocity, $B_{s} = \sqrt{\mu_{0}p_{s}} = 1.86$~G for magnetic field, $t_{s} = L_{s}/v_{s} = 5.42\times 10^{3}$~s for time, and $E_s = \rho_s v_s^2 L_s^3= 9.31\times 10^{31}$~erg for energy.

In the magnetic induction equation, the trigger of the magnetic
reconnection depends on the specific choice of magnetic diffusivity
$\eta$. To avoid this sensitivity issue, we use no explicit form of
$\eta$ in the magnetic induction equation,
following~\citet{jiang_FundamentalMechanismSolar_2021}. This approach
minimizes resistivity and maximizes the Lundquist number at given
spatial resolutions, as any non-zero $\eta$ would lead to greater
resistivity than the numerical resistivity alone. Consequently,
magnetic reconnection occurs only when the current layer becomes
sufficiently narrow, approaching the grid resolution, where numerical
diffusivity becomes significant.

\subsection{Grid settings}
The computational domain is a spherical shell ranging from the solar
surface to around $20$ solar radii, where the solar wind becomes
already supersonic and super-Alfv{\'e}nic. The lower boundary is set
at the solar surface, and more exactly, the base of the corona, while
the upper boundary is positioned far enough to study the near-Sun
propagation of a CME. We used a Yin-Yang grid to avoid the polar
problems (i.e., grid singularity) of the standard spherical
grid~\citep{jiang_NEWCODENONLINEAR_2012}. The Yin-Yang grid is
composed by two low latitude partial-sphere grids, identical but with
different orientations, to cover the full sphere with small patches
overlapped (see \Fig~1 of \citet{jiang_NEWCODENONLINEAR_2012}). The
computation utilizes block-structured adaptive mesh refinement (AMR),
which dynamically adjusts grid resolution based on the evolving
features during the simulation to improve accuracy and efficiency. For
this study, the base resolution in latitude ($\Delta \theta$) and
longitude ($\Delta \phi$) is set to $2^{\circ}$, and the AMR is
configured to achieve a maximum refinement level of $4$. Therefore,
the highest resolution is $0.25^{\circ}$. The grid cells are
configured to be close to regular cubes by setting
$\Delta r = r \Delta \theta$ (therefore not uniform in radial
direction), and the highest resolution near the solar surface is
$\Delta r \approx 4 \times 10^{-3}$. We pay particular attention to
the formation and reconnection of the current sheet in the
simulation. The formation and evolution of current sheet are tracked
by refining regions with both $J/B>4.0$ and plasma beta $\beta < 0.25$
to the highest level. Additionally, areas where
$|\nabla(B^2/2)|\Delta/\rho>15$ and $|(B\cdot\nabla)B|\Delta/\rho>15$
(where $\Delta$ is the grid resolution) are also refined to ensure
high resolution in regions with strong magnetic field gradients and
curvatures.

\subsection{Initial conditions}

The simulation is initialized with a plasma specified by the Parker's
classic spherically symmetrical model of solar wind and a potential
(i.e., current free) magnetic field. The Parker model is solved by
assuming an ideal adiabatic gas with $\gamma = 1.05$, and constrained
by a solar surface density $\rho = \rho_{s}$ and temperature
$T = 1.8~T_{s}$. The magnetic field comprises a background dipole
field to represent the global coronal magnetic structure during solar
minimum and an embedded small bipolar field representing an active
region. We first specify the magnetogram, i.e., a map of the radial
magnetic field component $B_r$ on the solar surface, and then compute
the corresponding potential field for the whole simulation volume,
using a fast solver~\citep{jiangUnifiedVeryFast2012}. The map is given by
$B_r = B_r^{\rm g} + B_r^{\rm a}$, with $B_r^{\rm g}$ for the global
field and $B_r^{\rm a}$ for the active region. The global component
$B_r^{\rm g}$ is assumed to the solar surface flux of a dipole
$\vec B_{\rm d}$ with magnetic moment $\vec{m} = (0,0,1.5)$ placed at
the solar center,
\begin{equation}
    \vec B_{\rm d} = \frac{3 (\vec{m}\cdot \vec{r}) \vec{r}}{r^5} - \frac{\vec{m}}{r^3}.
\end{equation}
The active region component is given by the sum of two 2D Gaussian
functions on the $(\theta,\phi)$ plane,
\begin{align}
	B_r^{\rm a} = & B_0 e^{-\frac{\theta'^{2}}{\sigma_\theta^2}} e^{-\frac{(\varphi' - \varphi_{\text{offset}})^2}{\sigma_\varphi^2}} - \notag \\
	& B_0 e^{-\frac{(\theta' + \theta_\text{slip})^2}{\sigma_\theta^2}}e^{-\frac{(\varphi' + \varphi_{\text{offset}})^2}{\sigma_\varphi^2}}.
\end{align}
Here $(\theta', \phi')$ are rotated coordinates with respect to the
original coordinates $(\theta,\phi)$ given by
 \begin{equation}
	\begin{pmatrix}
		\varphi' \\
		\theta'
	\end{pmatrix}
	=
	\begin{pmatrix}
		\cos \alpha & -\sin \alpha \\
		\sin \alpha & \cos \alpha
	\end{pmatrix}
	\begin{pmatrix}
		\varphi - \phi_0\ \\
		\theta - \theta_0
	\end{pmatrix}
\end{equation}
where $(\theta_0, \phi_0)$ is the center of the active region, and
$\alpha$ the rotation angle. By rotating the coordinate system, it is
convenient to mimic active regions with different
orientations. $\sigma_\theta$ and $\sigma_\phi$ control the extent of
the magnetic flux distribution in the $\theta'$ and $\phi'$
directions, respectively, and $\phi_{\rm offset}$ controls the
separation of the two polarities. Here the parameters are given as
$B_0 = 48$~G, $\theta_0 = 70^\circ$, $\sigma_\theta = 9^\circ$,
$\phi_0 = 180^\circ$, $\sigma_\phi = 9^\circ$,
$\phi_{\rm offset}=6^{\circ}$, and $\alpha = 45^{\circ}$. We further
used a parameter $\theta_{\rm slip}= 10^\circ$ to form a shearing
shape between the positive and negative polarities along the polarity
inversion line (PIL).
\Fig~\ref{fig:initialCondition} shows the $B_r$ map. As can
be seen, the field configuration resembles a typical bipolar solar
active region located in the northern hemisphere during it's decaying
phase, and sheared by the differential rotation of the Sun.

\subsection{Boundary conditions}

Our simulation consists of different stages featured by specifying
different velocity at the inner boundary (i.e., the solar
surface). One is a relaxation stage,
i.e., no external driver is applied, in which the surface velocity is simply
given as $v_r = v_{\theta} = v_{\phi} =
0$.  The other is a driving stage in which the surface velocity is
given as a surface rotation flow at each polarity of the AR to inject
free magnetic energy into the AR. Following~\citet{jiang_FundamentalMechanismSolar_2021}, the
driving flow is incompressible with streamlines aligning with the
contour lines of $B_r$, therefore not altering the profile of $B_r$ on
the surface. Specifically, the surface velocity is set as
\begin{equation}
  \label{velocity}
  v_{r}=0,
    v_\theta = \frac{1}{R \sin \theta} \frac{\partial \Psi(B_r)}{\partial \phi},
    v_\phi = -\frac{1}{R} \frac{\partial \Psi(B_r)}{\partial \theta}
\end{equation}
with $\Psi$ given by
\begin{equation}
    \Psi = k B_r^2 e^{-(B_r^2 - B_{r, \mathrm{max}}^2)/B_{r, \mathrm{max}}^2}
\end{equation}
where $B_{r, \mathrm{max}}$ is the maximum value of $B_r$ at the
surface, and $k$ is a scaling constant chosen so that the maximum
surface velocity is $25.6$~km~s$^{-1}$. The flow pattern is depicted
in \Fig~\ref{fig:initialCondition}. We note that this velocity is
about an order of magnitude higher than typical photospheric motion
speeds, which are approximately a few km~s$^{-1}$
\citep{amari_VeryFastOpening_1996,
  tokman_ThreedimensionalModelStructure_2002,
  torok_EvolutionTwistingCoronal_2003,
  devore_HomologousConfinedFilament_2008}. We intend to accelerate the
surface driving to compete the effects of numerical dissipation of the
accumulated free energy in the simulation, such that enough free
energy can be stored in the simulated active region to produce an
eruption. In~\citet{jiang_FundamentalMechanismSolar_2021}'s simulation
they used a very high resolution to reduce the numerical diffusion and
thus they were able to apply a rather slow driving speed of a few
km~s$^{-1}$ that is close to the actual photospheric motion. However,
for a global simulation in this paper, it is expensive to use a high
resolution to reduce the numerical diffusion, and therefore we choose
to enlarge the driving speed.

We fix the plasma density and temperature at the bottom surface to its
initial uniform value since the surface flow is incompressible. With
the velocity prescribed (either as zero or by \Eq~\ref{velocity}) and
$B_r$ unchanged, only the evolution of the horizontal magnetic field
needs to be solved, by using the magnetic induction equation,
\begin{equation}
  \frac{\partial \vec B}{\partial t} = -\nabla \times \vec E
\end{equation}
The equation is discretized using a 2nd-order difference in space and
a forward difference scheme in time. Specifically, we first compute
the electric field at the grid points by assuming that
$\vec E = - \vec v \times \vec B$, and then using a central different
in $\theta$ and $\phi$ directions and a one-sided 2nd-order difference
in $r$ direction. Taking the $B_\theta$ component as an example, the
induction equation is casted in spherical coordinates as
\begin{equation}
  \frac{\partial B_{\theta}}{\partial t} = \frac{1}{r\sin\theta} \frac{\partial
    (-E_{r})}{\partial \phi} -
   \frac{1}{r}\frac{\partial(-rE_{\phi})}{\partial r},
\end{equation}
for which the numerical scheme is given by
\begin{eqnarray}
  \frac{(B_\theta)_{i,j,0}^{n+1} - (B_\theta)_{i,j,0}^n}{\Delta t} =
  \frac{(-E_r)_{i,j+1,0}^{n+1/2} - (-E_r)_{i, j-1, 0}^{n+1/2}}{2r_{0}\sin{\theta_{i}} \Delta \phi}- \nonumber\\
  \frac{(\Delta r_{1}+\Delta r_{0})^{2}\Gamma_{1} -
  \Delta r_{0}^{2}\Gamma_{2} - \Delta r_{1}(\Delta r_{1}+2\Delta r_{0})\Gamma_{0}}
  {r_0\Delta r_{0} \Delta r_{1}(\Delta r_{0} + \Delta r_{1})}
\end{eqnarray}
where $\Gamma_{k} = r_{k}(-E_{\phi})_{i,j,k}$, the subscripts $i$,
$j$, $k$ represent the grid points in the $\theta$, $\phi$, $r$
directions respectively, with $k = 0$ corresponding to the points at
the bottom boundary (note that no ghost layer is used in our code),
and $\Delta r_{k} = r_{k+1}-r_{k}$. This approach allows for a
self-consistently update of the magnetic field and facilitates
simulation of the line-tied effect at the bottom boundary, which is
crucial for the success of this simulation. For the outer boundary, we
implemented non-reflecting conditions for all variables using the
projected-characteristic method~\citep[see details in,
e.g.,][]{hayashiMagnetohydrodynamicSimulationsSolar2005,
  wuDataDrivenMagnetohydrodynamic2006a,
  jiangRECONSTRUCTIONCORONALMAGNETIC2011,
  fengDATADRIVENMODELGLOBAL2012}.

\begin{figure*}
  \centering
  \includegraphics*[width=\textwidth]{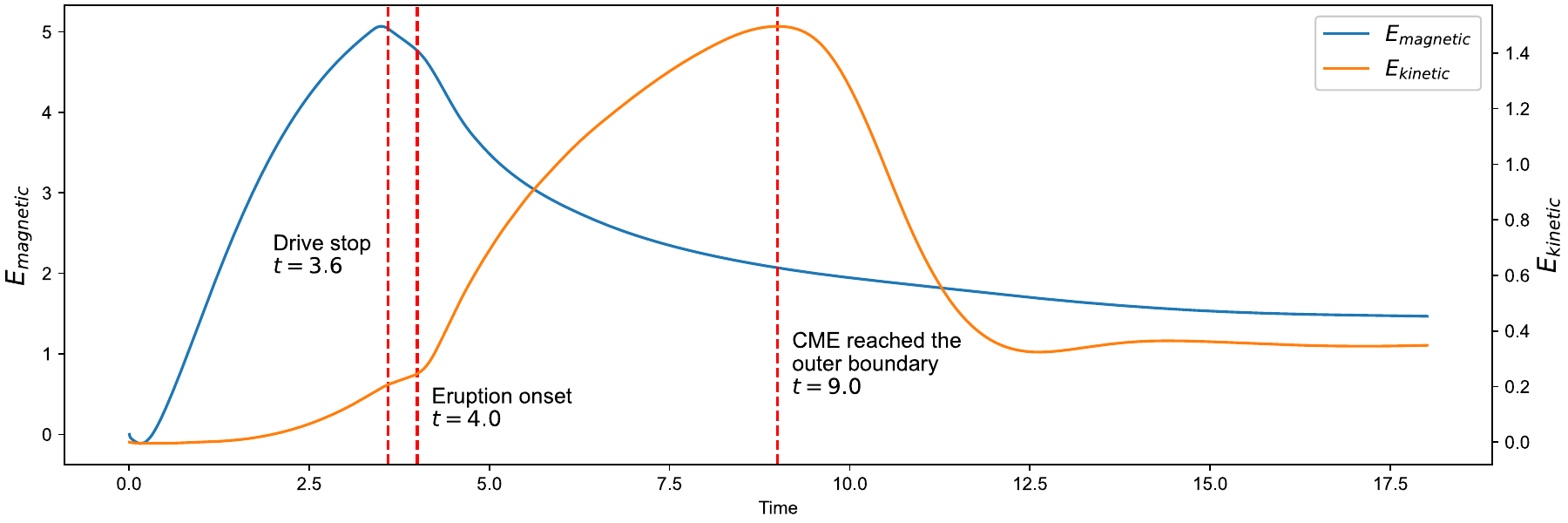}
  \caption{Evolution of magnetic energy (blue line) and kinetic energy
    (orange line) in the driving and eruption phases. The vertical dashed
    lines from left to right denote respectively the stopping time of surface
    driving, the eruption onset time, and the time when the leading
    edge of CME starts to leave the simulation volume.}
  \label{fig:energy evolution}
\end{figure*}

\begin{figure*}
  \centering \includegraphics[width=\textwidth]{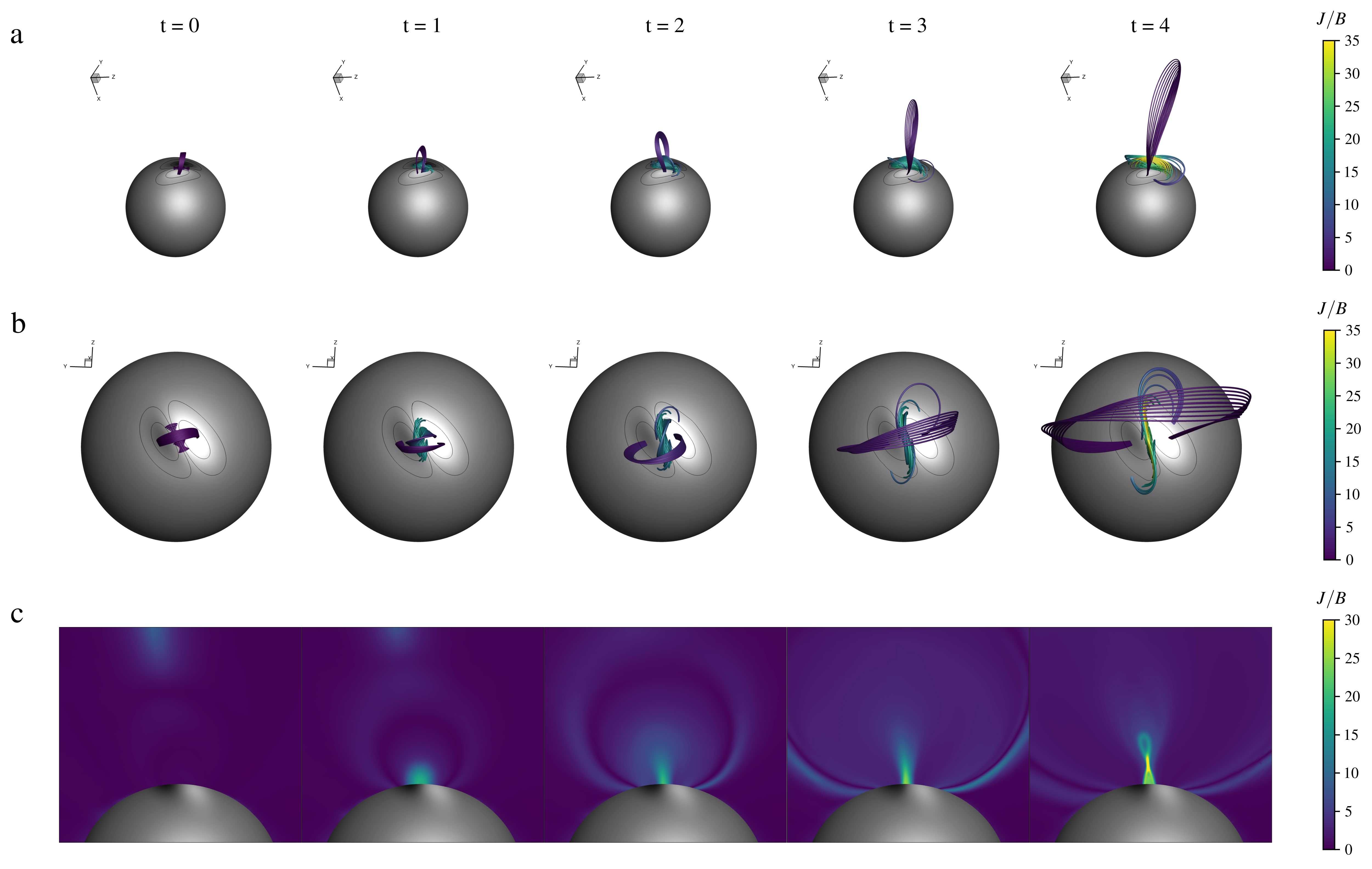}
  \caption{Evolution of magnetic field lines and electric currents
    prior to eruption. (a) Top view of magnetic field lines at
    different times in the simulation. The colored thick lines
    represent magnetic field lines, with colors denoting the value of the current density normalized by
    magnetic field strength ($J/B$). (b) 3D prospective
    view of the same field lines shown in panel a. (c)
    Vertical cross-section of $J/B$.}
  \label{fig:magnetic_evolution}
\end{figure*}

\begin{figure*}
	\centering
	\includegraphics[width=\textwidth]{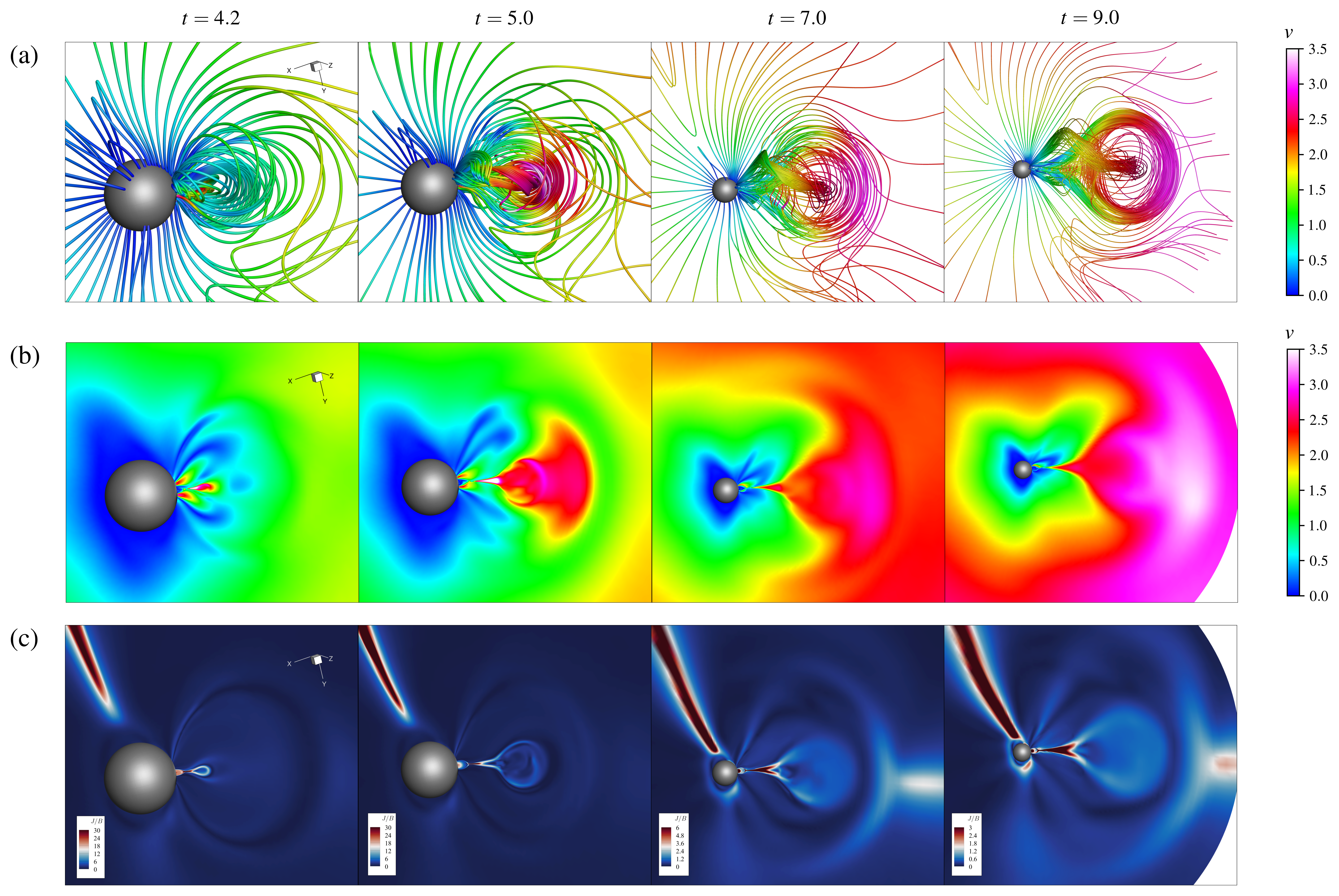}
	\caption{Evolution of magnetic field lines (a), velocity (b) and current density (c) of the CME. In all the panels, the gray sphere represents the surface of the Sun, and different fields of view are shown for different times. The magnetic field lines are pseudo-colored by velocity. The current density is shown normalized by the magnetic field strength (i.e., $J/B$). The velocity and current density are shown on the same cross section of the volume.}
	\label{fig:current and velocity evolution}
\end{figure*}

\begin{figure*}
    \centering
    \includegraphics[width=0.8\textwidth]{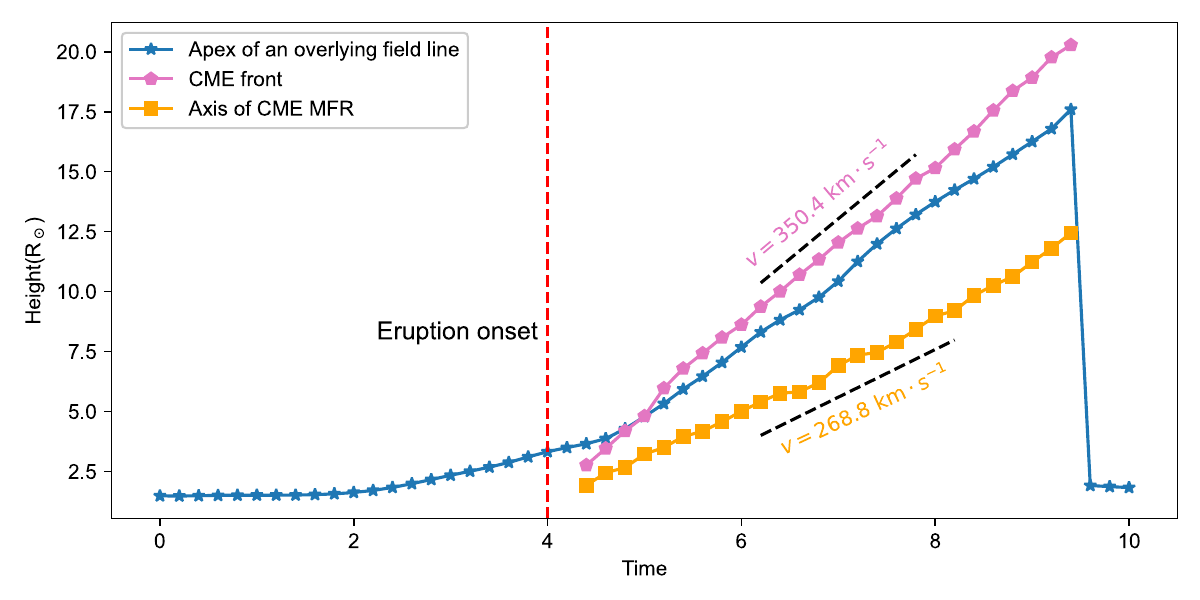}
    \caption{Variation of heights of the apex of an overlying magnetic field line of the active region's sheared core, the CME front and the axis of the CME MFR. The velocity of the structures is shown by the dashed lines with numbers.}
    \label{fig:kinematics}
\end{figure*}

\begin{figure*}
    \centering
    \includegraphics[width=\textwidth]{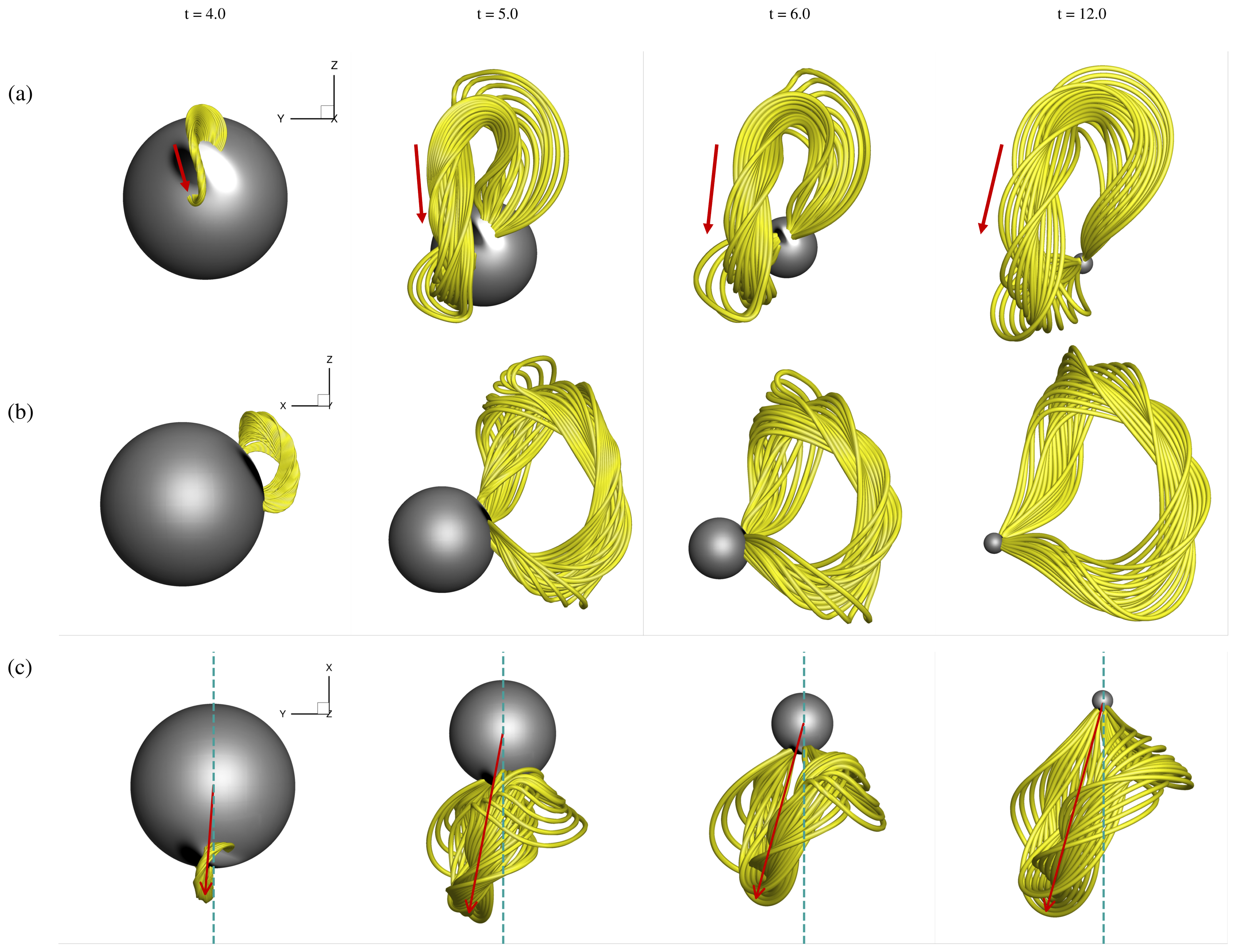}
    \caption{Evolution of the CME MFR at four different times (from left to right). Three view angles are shown from top to bottom. The sphere represents the solar surface shown with magnetic flux distribution. The thick lines show the magnetic field lines traced close to the axis of the MFR. In (a), the arrows show the direction of the axis of the MFR of CME at different times. In (c) the arrows the propagation direction of the MFR.}
    \label{fig:rotation}
\end{figure*}

\begin{figure*}
    \centering
    \includegraphics[width=\textwidth]{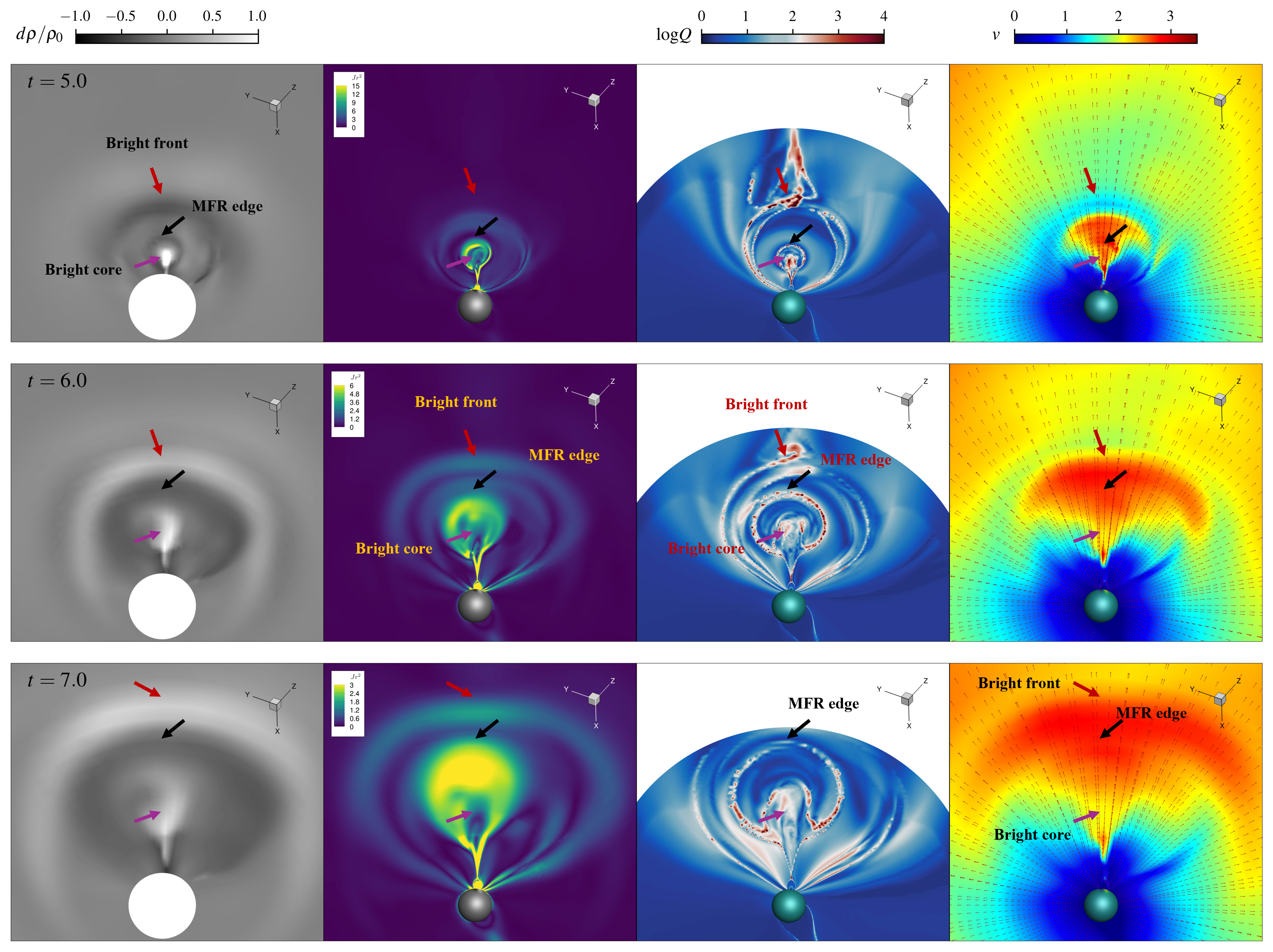}
    \caption{Cross section profiles of the density (the first column), current density (the second column), magnetic squashing factor (the third column) and the velocity (the last column) at three different times (from top to bottom). All panels show the same cross section of the volume with strictly the same view angle and field of view. For density, the ratio is defined as $d\rho/\rho_0 = [\rho(t)-\rho(0)]/\rho(0)$, and the area within 2 solar radii is blanked to mimic the occulter of coronagraph. For current density, a factor of $r^2$ is multiplied. The squashing factor is computed only up to 12 solar radii. For the velocity, the thin arrows show the direction of the flow. In all the panels, the three thick arrows, from top to bottom, denote respectively the location of the bright front, the edge of the MFR, and the bright core. Not that for comparison, the location of each arrow in the same time is identical in the four different panels.}
    \label{fig:threepart}
\end{figure*}

\section{Results}

We first perform a relaxation process (i.e., no surface driving flow)
to achieve a steady-state solar wind solution, which provides a
background for the subsequent simulation of CME triggering and
evolution. \Fig~\ref{fig:backgroundSolarWind} shows the relaxed
magnetic field lines and the evolution of energies during the
relaxation process. As driven by the solar wind, the magnetic field
lines from the two poles become eventually open, while at the low
latitudes forms the helmet-like coronal streamer. With opening of the
field lines, the magnetic energy is increased at the price of the
kinetic energy loss. The relaxation process is stopped at $t=40$, when
both the two energies become almost unchanged (albeit that the
magnetic energy shows slow decrease due to the numerical
resistivity). Once this relaxed state is established, we then
introduce the rotational velocity to the active region to simulate
photospheric motion that injects free magnetic energy into the system.

\Fig~\ref{fig:energy evolution} shows the energy evolution, and the
time is reset with $t=0$ for the start of applying the surface driving
flow. Note that the background values are subtracted from the
energies, and therefore at $t=0$ both the energies are zero. As can be
seen, with the driving flow applied, the magnetic energy keeps
increasing. The kinetic energy also increased but is much slower. This
indicates that the AR system evolves mostly quasi-statically. Since
the driving speed is much higher than the real photospheric speed, to
avoid a too much twisting of the field lines, we turn off the surface
driving at $t=3.6$, shortly before the eruption onset time of
$t=4.0$. At $t=4.0$, i.e., onset of the eruption, the kinetic energy
shows a rapid increase and the magnetic energy shows a rapid
decrease. This transition corresponds to fast release of the free
magnetic energy, which drives an impulsive acceleration of the plasma.

\Figs~\ref{fig:magnetic_evolution} and \ref{fig:current and velocity
  evolution} illustrate the evolution of magnetic field lines and
current density structures before and during the eruption. The
evolution is almost identical to that
of~\citet{jiang_FundamentalMechanismSolar_2021}'s simulation in a
local Cartesian coordinates. The pre-eruption stage is featured by a
slow shearing and expansion of the active region field within which a
current sheet is gradually formed above the PIL. Due to the maximum
gradient of velocity along the PIL, strong shear develops in the
magnetic field lines at that location, forming an S-shaped
structure. This S-shape is essentially composed of two sets of
J-shaped magnetic field lines, which are created as the magnetic field
in the active region is sheared in opposite directions by the
rotational flow. Initially, the current is distributed volumetrically,
but it is subsequently compressed into a vertical, narrow layer
extending above the PIL, i.e., the current sheet
(\Fig~\ref{fig:magnetic_evolution}c).

The eruption is triggered once magnetic reconnection starts in the
current sheet. An MFR originates from the tip of the current sheet and
quickly ascends, leaving behind a cusp-shaped structure that separates
the post-flare loops from the un-reconnected magnetic field regions
(\Fig~\ref{fig:current and velocity evolution}). As driven by the
ongoing reconnection, the MFR rapidly grows, forming a
CME. \Fig~\ref{fig:kinematics} shows the time-height profile of the
CME, including the leading edge of the CME and the apex of the MFR
axis. As can be seen, the leading edge of the CME shows a radial
velocity of approximately $350$~km~s$^{-1}$, and the rope axis has a
velocity of $270$~km~s$^{-1}$. In addition, we traced the motion of
the field line with footpoint fixed at the center of the positive
polarity of the AR. This field line can correspond to a coronal loop
in observation, initially located in the core of the active
region. The coronal loop rises slowly from $t=2$ to around $4.5$,
resembling the slow-rise phase in observed initiation process of many
CMEs~\citep{zhang_StatisticalStudyMain_2006,
  cheng_InitiationEarlyKinematic_2020}.  The kinetic energy gained by
the plasma, primarily in the CME, accounts for approximately one-third
of the released magnetic energy
(\Fig~\ref{fig:magnetic_evolution}). This suggests that the remaining
two-thirds of the energy is consumed by the flare, which aligns with
the typical energy partitioning between flares and CMEs in eruptive
events \citep{emslie_GlobalEnergeticsThirtyeight_2012}. At around
$t=9$, the leading edge of CME reaches the outer boundary, and with
the leaving of the CME from the computational volume, the kinetic
energy decreases and is eventually restored to its pre-eruption value.

\Fig~\ref{fig:rotation} shows the evolution of the MFR, for which
sampled magnetic field lines near the axis are plotted. Besides the
growth (due to reconnection) and expansion, the MFR experiences a
noticeable rotation that is often observed in the early phase of CME
evolution~\citep{manchester_PhysicalProcessesCME_2017}. At the very
beginning of the eruption, the MFR axis directs mainly along the PIL
of the AR (see the panels for $t=4$) since the MFR is formed from
reconnection of the highly sheared arcade. Subsequently, the MFR
rotates clockwise into a mainly southward direction. If direct to the
Earth, this simulated CME could render a strong geomagnetic effect as
it has a southward magnetic field component. The direction of rotation
is consistent with the findings
in~\citet{zhouRelationshipChiralitySense2020a} that during its
eruption an forward (reverse) S-shaped MFR, as manifested by eruptive
filament, rotates clockwise (anti-clockwise), and can be explained by
the interaction of the erupting MFR with the background
field~\citep{zhouMechanismMagneticFlux2023}. The MFR is also deflected
to the east, although slightly, during its propagation, which can be
seen in the bottom panels of \Fig~\ref{fig:rotation}. It is known that
the deflection of CME can be attributed to the interaction of CME with
the solar wind~\citep{wangDeflectionCoronalMass2004}, e.g., a CME
faster (slower) than the ambient solar wind would be deflected to the
east (west). Our simulation supports this, as here the CME has a speed
larger than that of the ambient solar wind during the simulated time
interval (see \Fig~\ref{fig:current and velocity evolution}). The
deflection of CME is also an important factor in determining the
geomagnetic effect by changing the propagation path.

\Fig~\ref{fig:threepart} (left column) shows the density profile on a
central cross section of the CME. The cross section is roughly
perpendicular to axis of the MFR. We used the ratio
$d\rho/\rho_0 = [\rho(t)-\rho(0)]/\rho(0)$ to highlight the variation
of the density relative to its background value. As can be seen, the
density profile presents a three-part structure, namely a bright core,
a dark cavity and a bright front, which resembles typical coronagraph
observations. To understand the relationship between the three parts
and the corresponding magnetic configuration, we plot the distribution
of current density in the second column of
\Fig~\ref{fig:threepart}. Due to the fast expansion of the MFR, the
current density is multiplied by a factor of $r^2$ to more clearly
show the entire structure of the MFR. For more precisely locating the
interface between the MFR and the overlying field, we also computed
the magnetic squashing factor~\citep{1999A&A...351..707T,
  demoulinExtendingConceptSeparatrices2006}, from which the
quasi-separatrix layer corresponding to the boundary of the MFR can be
identified. Compared in the right column of \Fig~\ref{fig:threepart}
is the distribution of velocity on the cross section, which is useful
to analyze how the different structures of the density is formed.
Since there is no filament in our simulation, the bright core does not
correspond to a filament. By comparing the density profile and the
current distribution, we can see that the bright core is located
mainly at the lower part of the erupting MFR. Furthermore, the
formation of the bright core can be understood from structure of the
velocity; the fast reconnection outflow, driven by the strong
slingshot effect (i.e., the outward magnetic tension force) of the
newly-reconnected field lines, injects continuously plasma into the
MFR, and this plasma is piled up at the lower part of the MFR because
the fast reconnection jet is decelerated briefly after merging into
the MFR~\citep[see also][]{jiang_FundamentalMechanismSolar_2021}. As a
result, the density becomes high there, forming the bright core. The
dark cavity initially corresponds to the weakly sheared field
overlying the highly-sheared core. During subsequent evolution, this
field gradually reconnects and joins into the MFR as its envelope
part, and thus part of the dark cavity now corresponds to this
envelope part of the MFR. The low density in the cavity is a result of
the fast expansion of the magnetic flux as it moves out and the
ambient magnetic pressure decreases. Ahead of the dark cavity is the
bright front, which is formed due to the compression of the plasma
ahead of the CME's high-speed region.

\section{Conclusions}

In this study, we presented a global-corona MHD simulation of the
formation of a CME and its interaction with the ambient solar wind. We
first constructed a background solar wind by relaxing the Parker's
solution with a global dipole field, which embeds a local bipolar
field that represents an active region, to an MHD equilibrium. Then we
energized the active region field using continuous shearing motion
along the PIL until an eruption is produced. Our simulation
encompassed the entire process from the gradual accumulation of
magnetic energy to the catastrophic release of magnetic free energy
that initiates the CME. The mechanism of CME initiation is in line
with what has been shown in a previous simulation casted in local
Cartesian coordinates~\citep{jiang_FundamentalMechanismSolar_2021}; an
internal current sheet gradually forms within a sheared magnetic
arcade and fast reconnection at this current sheet triggers and drives
the eruption.

We further analyzed the subsequent evolution and propagation of the
CME to around $0.1$~AU, highlighting key aspects such as the formation
of MFR and its kinematic characteristics, deflection, rotation, and
morphology, which may shed light on interpretation of
observations. The MFR is originated and grows from the ongoing
reconnection in the current sheet. Its axis reaches a speed of around
$270$~km~s$^{-1}$, while the CME front has a speed of
$350$~km~s$^{-1}$, somewhat faster than the simulated solar wind. As a
forward S shaped MFR, it rotates clockwise during the evolution, and
also exhibits a eastward deflection by interacting with the ambient
solar wind. From the cross section profile of plasma density, the CME
exhibits a typical three-part configuration. The bright core is mainly
located at the lower part of the MFR and is produced by the plasma
that is first rapidly ejected by the high-speed reconnection outflow
and then piled up at the lower part of the MFR due to the brake down
of the reconnection outflow. Therefore, a CME owning a bright core
does not necessarily to contain a erupting filament, but with a
filament, the core should be of course more prominent. The dark cavity
contains both outer layer of the MFR and its overlying field (which is
gradually integrated into the MFR due to the reconnection) that
expands rapidly as the whole magnetic structure moves out. The bright
front is formed due to the compression of the plasma ahead of the
fast-moving magnetic structure.

We note that our model is still far from a realistic description of
CME initiation and evolution. Future developments are needed, for
example, by constructing a more realistic background solar wind with
empirical coronal heating and acceleration that can produce a two-mode
(i.e., fast and slow) wind structure~\citep{fengHYBRIDSOLARWIND2011};
the observed synoptic magnetograms should be used to construct the
global coronal magnetic field such that the CME can be initiated in a
realistic magnetic environment~\citep{mikicPredictingCorona212018,
  torokSuntoEarthMHDSimulation2018}; furthermore, the data-driven
technique based on vector magnetograms should be used to follow the
evolution and eruption of real active
regions~\citep{jiangDatadrivenModelingSolar2022}. With these
improvements, a in-depth understanding of the birth, 3D structure and
evolution of CME is hopeful to achieve, and the model can be
incorporated into Sun-to-Earth space weather modelling framework.



\section*{Acknowledgements}
This work is jointly supported by National Natural Science Foundation
of China (NSFC 42174200), Shenzhen Science and Technology Program
(Grant No. RCJC20210609104422048), Shenzhen Key Laboratory Launching
Project (No. ZDSYS20210702140800001), Guangdong Basic and Applied
Basic Research Foundation (2023B1515040021).
\section*{Data Availability}
All the data generated for this paper are available from the authors
upon request.


\bsp	
\label{lastpage}
\end{document}